\documentclass[intlimits,twoside,a4paper]{article}

\usepackage[cp1251]{inputenc}

\usepackage[eqsecnum]{cmpj3}

\newcommand{\be}{\begin{equation}}
\newcommand{\ee}{\end{equation}}
\newcommand{\bea}{\begin{eqnarray}}
\newcommand{\eea}{\end{eqnarray}}
\newcommand{\eqan}[1]{\begin{eqnarray*}#1\end{eqnarray}}

\def \do{\Delta_{\cal O}} 

\issue{2023}{26}{4}{43605}
\doinumber{10.5488/CMP.26.43605}

\title[Entropic force in ring polymer solutions with different topologies in a slit]%
{Entropic force in a dilute solution of real ring polymer chains with different topological structures
 in a slit of two parallel walls with mixed boundary conditions%
\thanks{Former name Zoryana Usatenko}}

\author[P. Kuterba, Z. Danel, W. Janke]{P. Kuterba\orcid{0000-0002-3442-5854}\refaddr{label1,label2}, 
Z. Danel\orcid{0000-0003-0147-5422}\refaddr{label3}\thanks{Corresponding author:\email{zoriana.danel@pk.edu.pl}},
W. Janke\orcid{0000-0002-5165-9097}\refaddr{label4}}
\addresses{
\addr{label1} Institute of Theoretical Physics, Jagiellonian University,  prof. S. {\L}ojasiewicza St. 11, 30--348 Cracow, Poland
\addr{label2} Faculty of Medicine, Jagiellonian University Medical College, {\L}azarza St. 16, 31--530 Cracow, Poland
\addr{label3} Faculty of Materials Engineering and Physics, Cracow University of Technology, Podchora\.{z}ych St.  1 , 30--084 Cracow, Poland
\addr{label4} Institut f\"ur Theoretische Physik, Universit\"at Leipzig, IPF 231101, 04081 Leipzig, Germany
}

\date{Received September 5, 2023, in final form November 21, 2023}
\begin{document}
\maketitle
\begin{abstract}
 The molecular dynamics simulations were used to obtain the radius of gyration of real ring polymer chains with
different topological structures consisting of 360 monomers.
 We focus on the entropic force which is exerted by a dilute solution of ring polymer
chains of different topological structures with the excluded volume
interaction (EVI) in a good solvent on the confining parallel walls of a slit  geometry. We consider
mixed boundary conditions of one repulsive
wall and the other one at the adsorption threshold. 
The obtained molecular dynamics simulation results for a wide slit region
demonstrate a qualitative agreement with previous analytical results
for ideal ring polymers. These results could lead to interesting potential
applications in materials engineering and improve understanding of
some biological processes suggested in the paper.  Additionally, they
could be applied in micro- and nano-electromechanical devices
(MEMS and NEMS) in order to reduce the static friction.
\printkeywords
\end{abstract}

\section{Introduction}
\setcounter{equation}{0}
Biopolymers such as deoxyribonucleic acid (DNA)  can have  in
nature a ring topology. One example is the Escherichia coli (\textit{E. coli}) bacteria, which is resistant to antibiotics.
Additionally,
the DNA of some viruses, such as bacteriophages $\lambda$ that attack
and destroy \textit{E. coli} bacteria can oscillate between linear and ring
topology as it was mentioned in
\cite{BergTymoczkoStryer02,ArsuagaVazquezTriguerosSumnersRoca02}.
The DNA structure might depend on whether it is inside or outside the
host cell. In the first case, it can switch to ring, in the other case it
remains linear in mature viruses. 
{  
Understanding those processes might
help in the development of phage therapy. 
This therapy serves as an alternative to antibiotic treatments, especially for antibiotic-resistant bacteria.}
{ Besides,  the
measurements of the depletion force between repulsive
wall and a single colloidal particle immersed in a dilute solution
of nonionic linear polymer chains in a good solvent were performed in a series of papers~\cite{RBL98,VCLY98,OSO97}. 
}
The depletion force enters  a complex force balance with other
intercolloidal interactions such as: electrostatic repulsion and van
der Waals attraction proposed by Derjaguin, Landau, Verwey, Overbeek
(DLVO) or hydrophobic interaction and hydration
\cite{cosgrove_colloid_2010,lekkerkerker_colloids_2011,curtis_depletion_2015}.

Polymers are large macromolecules, consisting of a huge number $N$
of monomers with $N\to\infty$, which can be well described by
statistical physics methods. In 1972, de Gennes \cite{G72}
introduced a concept of the polymer-magnet analogy for polymers in
infinite space, which was later extended by Barber et~al.~\cite{B78}
for polymers in confined geometries like semi-infinite space with
surface.

In a recent paper \cite{JHZU},
the monomer density profiles of a dilute solution of ideal ring
polymers in a $\Theta$-solvent between two parallel walls were
investigated under three different sets of boundary conditions (b.c.):
Dirichlet-Dirichlet (D-D), Dirichlet-Neumann (D-N) and
Neumann-Neumann (N-N).
{As demonstrated in \cite{UH17}, a dilute solution of ideal ring polymers with D-D b.c. exhibits an attractive depletion force, similar to linear polymer chains \cite{RU09}.}

Besides, the calculations of the
dimensionless depletion interaction potential and the depletion
force for the case of a dilute solution of real ring polymers with
the excluded volume interaction (EVI) in a good solvent for the case of D-D b.c. were obtained
\cite{UH17}. However, in \cite{UH17} it was reported that in a dilute
solution of ideal ring polymers with mixed D-N b.c., which corresponds
to the situation of one repulsive and the other wall at the
adsorption threshold, the respective depletion force starts to be
repulsive similarly to the case of a dilute solution of real linear
polymer chains with the EVI in a slit with N-N b.c.~\cite{RU09}. 
{One of the examples 
of such systems with mixed D-N b.c. can be solution of colloidal particles, nanoparticles or ceramic nanopowders with different adsorbing
and repelling properties in respect to polymers. Another example of such boundary conditions can be observed when a solution of 
colloidal or nanoparticles displays an opposite interaction with the polymers than the confining walls of reservoir where this polymer solution is inserted. }

Unfortunately, in \cite{UH17} only the case of ideal ring polymer chains 
was discussed analytically for the case of mixed b.c. In order to verify our analytical results in the
present paper we performed molecular dynamics simulation of a dilute
solution of ring polymer chains with the EVI in a good solvent
confined in a slit geometry of two parallel walls with mixed
boundary conditions. We discuss calculations of the radius of gyration and entropic
force, which is exerted by ring polymers with different topological
structures on the confining walls.

It should be mentioned that the depletion force for linear polymer
chains was studied previously with various numerical methods
\cite{MB98,MB98ol,HG04}. Shorter polymer chains ($32\leqslant N \leqslant 512 $) were
modelled using computationally expensive off-lattice Monte Carlo
method~\cite{MB98,MB98ol}. The problem of the modelled short chains was overcome
with the on-lattice  pruned-enriched Rosenbluth method (PERM), which
was used in three spatial dimensions to study the systems of long
flexible linear polymer chains ($N\leqslant 80000$) trapped between two
parallel walls \cite{HG04}. The ideal polymer chains were modelled as
random walks (RW) and real polymers with the EVI in a good solvent as
self-avoiding-walks (SAW). The depletion effects were studied in \cite{TMCM06} for
self-avoiding polymers within the framework of Langevin dynamics, where
they  discussed the competition between depletion attraction and
enhanced viscous friction as two competitive crowding-induced
effects.

The statistical mechanical properties of ring polymer chains in a slit with D-D b.c. were
investigated in a series of papers
\cite{JvanRensburg07,MatthewsLouisYeomans09}. As it was shown in
\cite{JvanRensburg07,MatthewsLouisYeomans09}, when the complexity
and compactness of ring polymers increases, then the radius of gyration
decreases at the same time. It makes ring polymers less mobile in the
systems without driving forces, as it was shown in
\cite{B20}. It should be noted that in the limit $N \to \infty$, the
influence of knots on polymer properties is much smaller. Thus, the
average size of ring polymers is proportional to $N^{\nu} $, where
$\nu$ is the Flory exponent.

The calculations of the entropic forces in a dilute solution of ring
polymers under confinement of two parallel repulsive walls (D-D
b.c.) obtained with the molecular dynamics simulations in
\cite{MatthewsLouisYeomans09} showed that 
polymers with more complex knots in the narrow slits produce higher entropic forces
exerted on the wall than the unknotted knots. The opposite relation is
observed  for relatively wide slits. Significant differences in
free energy and in metric properties between freely jointed polymers
and unknotted ring polymers were discussed in \cite{LiSunAnWang15}. It
was shown \cite{LiSunAnWang15} that the polymers with knots might
present different qualitative properties due to steric effects. The
knotting probability as it was shown in \cite{ED17}  using the method of
self-avoiding polygons depends on the polymer length and on the
magnitude of the excluded volume effects. Taking into consideration
the complexity of biological  systems one could expect that the
diversity of the structures and types of polymers could induce the
creation of some knots. It was shown that nematic ordering
(effective stiffening of DNA) induces the formation of torus knots
in phage capsids~\cite{RCSV2012}.

{In biological eukaryotic organisms, chromosomes separate during cell division. 
Centrosomes, primarily composed of centrioles, organize the structure of these dividing cells.}
So, centrioles are key ingredients in the proper separation process of the
parent cell into the child cells. 
However, there are species that lack centrioles, and a remarkable example being mutant flies (Drosophila
melanogaster) that do not have these structures and yet can normally develop until a certain life stage
\cite{BL06,DL20}. This means that despite the lack of centrioles, the
cells have successfully divided a great number of times. Centrioles are abundant
organelle in biological cells except a few taxonomic groups such as
coniferes (Pinophyta). One could imagine that depletion effects,
which create additional forces in constrained systems could play
some role in this or in similar biological mechanisms.

{ 
 In mammalian cells, protein synthesis on the rough endoplasmic reticulum (ER) results in proteins either passing into the ER lumen, interweaving through the membrane, or becoming transmembrane proteins.
The flat shape of the ER locally resembles a slit space, akin to a polymer in a slit geometry, where protein is attached to one side of it.
One of the examples of translationally active cells are liver cells. 
When analyzing the biological cells, we can find direct analogies to the behaviour
of polymers in a slit geometry between two parallel walls.
Ribosomes on the ER surface have some density so that they are spaced approximately at a distance between two membranes in ER sheets, approximating the situation of a dilute polymer solution between two parallel walls with mixed b.c. with one repulsive and one attractive wall (D-N b.c.).
Considering the emerging proteins can form various loops or knots in their structure due to their length. They can transfer interactions from one ER wall to the other one, reinforcing the analogy of proteins as polymers in a slit geometry.
}

\section{The model}
The present paper is devoted to the investigation of a dilute
solution of real ring polymers with the EVI in a good solvent
confined in a slit geometry with two parallel walls with mixed b.c. In
a dilute solution, different polymer chains do not overlap, so such
polymer solution can be well described by the statistics of a single
polymer chain. The polymer chain is confined in the slit between
parallel walls at  $z=0$ and  $z=L$. The walls are impenetrable, so the
potential tends to infinity $U \to \infty$ at walls. The short-range attractive potential is different from zero only over a
distance comparable with the effective monomer linear dimension
$\tilde{l}$. The single ideal polymer chain in a $\Theta$-solvent
can be modelled by random walk (RW) with the Flory critical exponent
$\nu=0.5$. The real polymer chain with the EVI in a good solvent for
temperatures $T$ above the $\Theta$-point can be modelled by
self-avoiding walk (SAW) with the Flory exponent $\nu=0.588$.

We performed molecular dynamics (MD) simulation using LAMMPS
(Large-scale Atomic Molecular Massively Parallel Simulator) for ring
polymers consisting of $360$ monomers. The neighbouring beads
interactions were modelled with FENE potential, the monomer repulsion
was modelled with 12-6 Lennard-Jones (LJ) potential and the
monomer-wall interaction with 9-3 LJ potential \cite{AS77}. We 
used Verlet integration algorithm with $\Delta t=0.005$ and the
Langevin thermostat, which helped us to keep temperature $T=1$
constant. Simulations were equilibrated for $10^7$ steps and the
simulation data collected over $2 \cdot 10^7$ steps. The results for
the force  were taken from 10 separate simulations for each of the slit widths used. 
Additionally, based on a similar system, we calculated the	radius
of gyrations for
each of the knot topologies. The calculations were performed for the
adsorption temperature $T=2.30$~\cite{KremmerBinder}. We  used
a model similar  to \cite{KCDJ23}. The interaction between the non-neighbouring monomers
was modelled with the 12-6 Lennard-Jones potential: 
\bea U_{\rm LJ}(r)=4 \epsilon
\left[ \bigg( \frac{\sigma}{r}\bigg)^{12} -\bigg(
\frac{\sigma}{r}\bigg)^{6}  \right], \eea for   $r \leqslant
r_{\rm min} = 2^{\frac{1}{6}} \sigma$ with $\sigma = 0.7/2^{\frac{1}{6}}= 0.6236$, where $r_{\rm min}$ is the extremum obtained from $\rd U_{\rm LJ} (r) /\rd r = 0$. 
The potential was truncated $U_{\rm LJ} (r) = 0$ for  $r>r_{\rm min}$ and shifted with $U_{\rm LJ, sh} (r) = U_{\rm LJ} (r) - U_{\rm LJ} (r_{\rm min}) $ 
in order to get $U_{\rm LJ, sh} (r_{\rm min}) = 0$.
The interaction
between monomers and walls was modelled with Lennard-Jones 9-3
potential \cite{AS77}: \bea U_{\rm LJ}^{\rm wall}(z)= \epsilon_w \left[ \frac{2}{15}
\bigg( \frac{\sigma}{z}\bigg)^{9} -\bigg( \frac{\sigma}{z}\bigg)^{3}
\right], \eea where $\epsilon_w$ is equal 1.0,  $z \leqslant
z_c$ setting  $z_c = z_{\text{min}}$ being obtained from the condition $\rd U^{\rm wall}_{\rm LJ}/\rd r =0$. 
So $z_{\rm min} =  (2/5)^{1/6} \sigma = 0.5353\sigma$  
for the repulsive wall and for the attractive wall, the potential was truncated at $z_c = 4.5
\sigma$. Shifting and truncation were performed for those potentials according 
to $U_{\rm LJ,sh}^{\rm wall} (z) = U^{\rm wall}_{\rm LJ}(z) - U^{\rm wall}_{\rm LJ} (z_c)$.
The
bonds were simulated with Finitely Extensive Nonlinear Elastic
(FENE) potential~\cite{KG90}: 
\bea U_{\rm FENE}(r)=-\frac{K R_0^2}{2}
\ln\left[ 1 - \bigg( \frac{r}{R_0}\bigg)^2 \right], \quad
{\tiny r <  R_0}, 
\eea 
where $R_0=1.0$ and $K=40.0$, while for $r>R_0$ the potential
$U_{\rm FENE}(r)=0$. We   performed simulations for linear, ring
polymers: $3_1,6_1,7_1,9_1,10_{124}$ and twist knots with $n=20$
number of twists.

\begin{table}[!b]
	\caption{The radius of gyration for different types of knottings
		obtained by molecular dynamics simulations.}
	\vspace{2mm}
	\centering
	\begin{tabular}{ccc}\hline\noalign{\smallskip}
		Knot type & Illustrative image & Radius of gyration \\  \noalign{\smallskip}\hline\noalign{\smallskip}
		$3_1$   & \includegraphics[width=0.8cm]{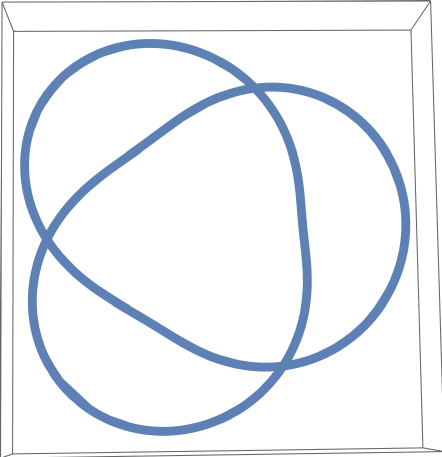} & 4.0893 \\
		$6_1$   &  \includegraphics[width=0.8cm]{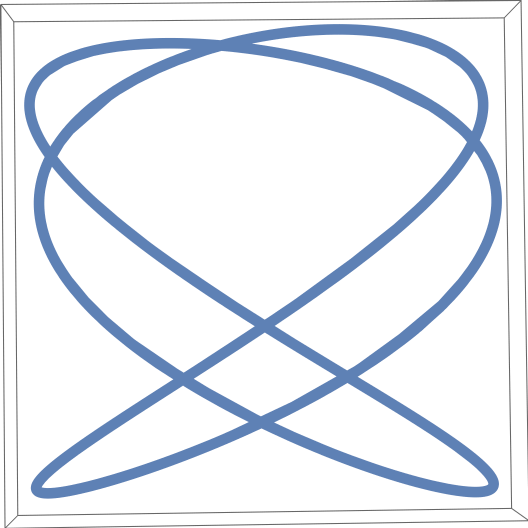} & 3.5128\\
		$7_1$   & \includegraphics[width=0.8cm]{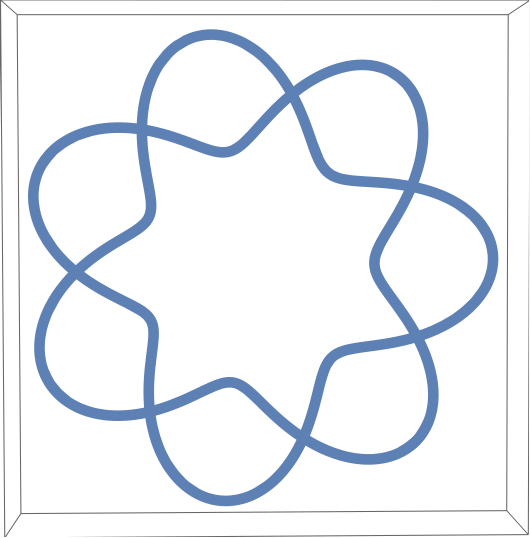} & 3.4318\\
		$9_1$   & \includegraphics[width=0.8cm]{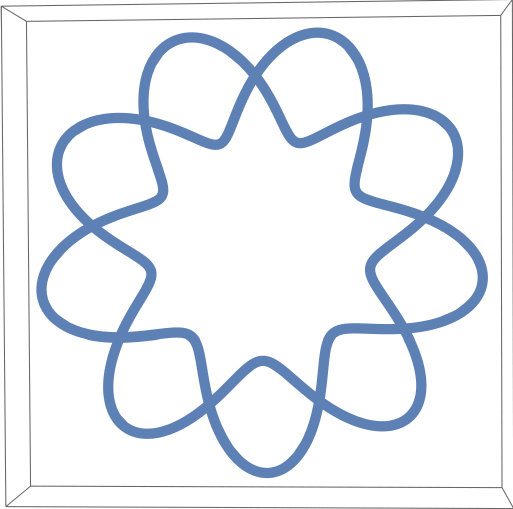} & 3.2817\\
		$10_{124}$ & \includegraphics[width=0.8cm]{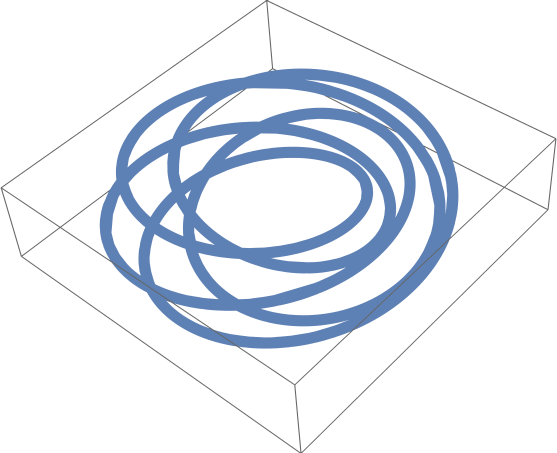} & 3.2355\\
		Twist knot with $n$=20 &\includegraphics[width=0.8cm]{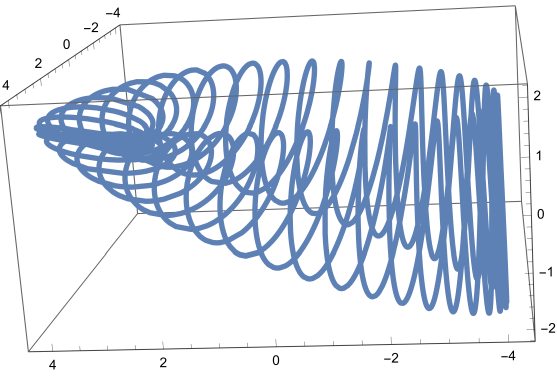} & 2.8073 \\\noalign{\smallskip}\hline
	\end{tabular}
	\label{knotrg}
\end{table}

\section{Results}

\begin{figure}[!t]
	\begin{center}
		\includegraphics[width=10.0cm]{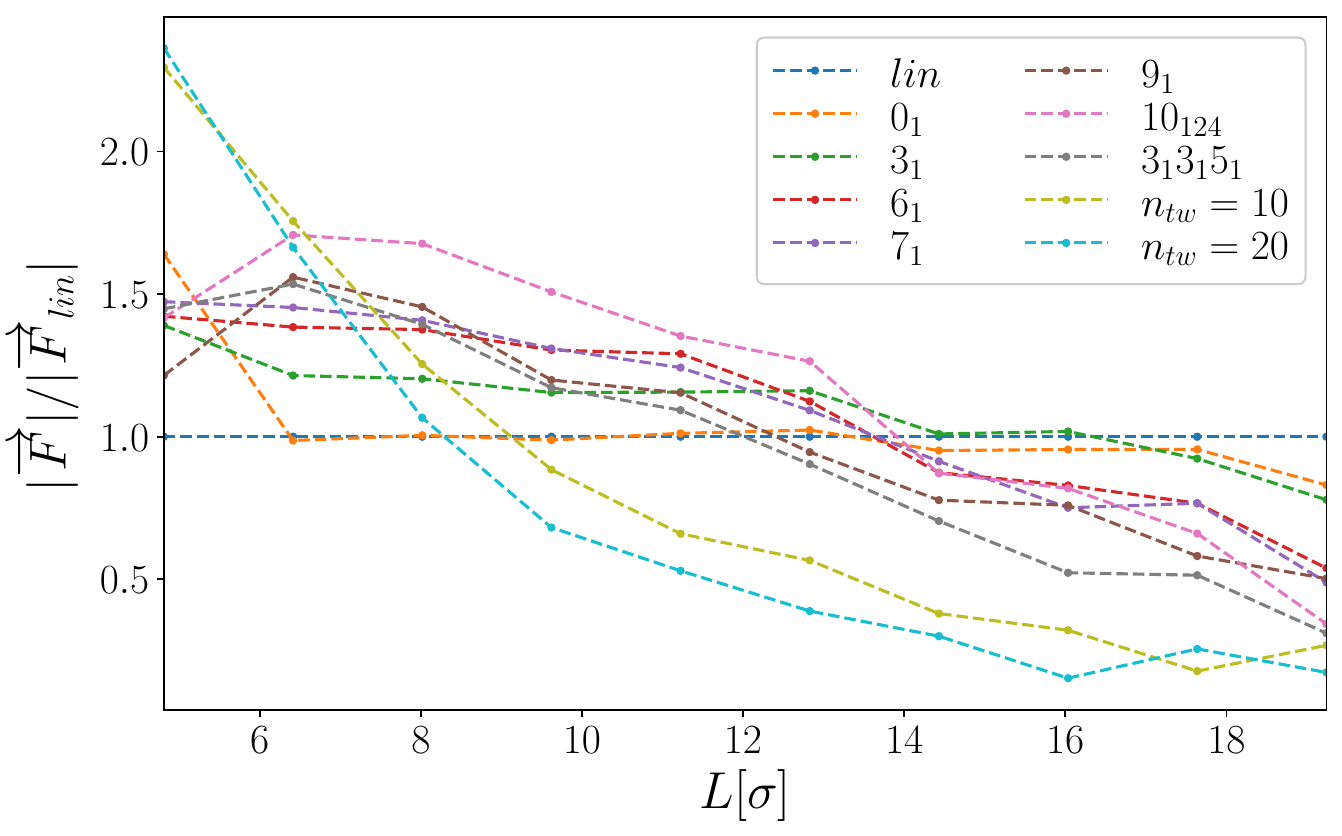}
		\caption{(Colour online) The ratio of entropic forces for a single real ring polymer chain to the force for real linear polymer chain in a
			slit geometry with one adsorbing wall and the other repulsive one
			obtained by molecular dynamic simulation.}
		\label{fig:fratio}
	\end{center}
\end{figure}

\begin{figure}[!t]
	\begin{center}
		\includegraphics[width=12cm,height=7cm]{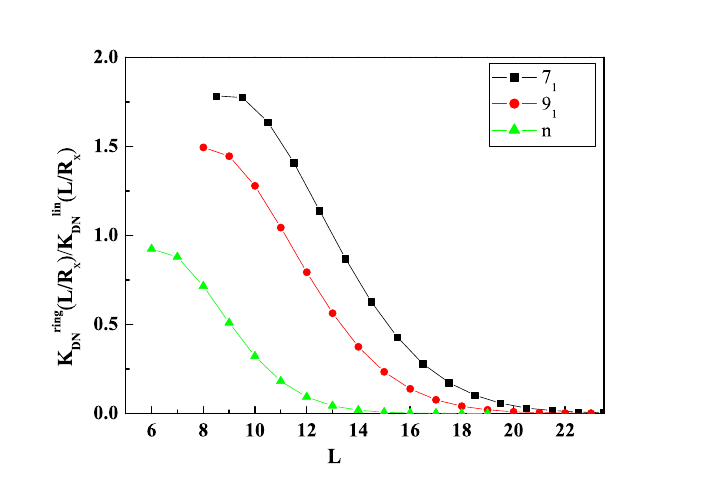}
		\caption{(Colour online) The entropic force ratio of specific knotted ideal ring polymers ($7_1$, $9_1$, $n_{ \rm tw} = 20$) to the linear ones obtained analytically on the 
			basis of the analytical results presented in  \cite{UH17,RU09} for a wide slit region. 
		}
		\label{fig:force}
	\end{center}
\end{figure}

Taking into account the dependence of the radius of gyration on the
topological structure, we investigated the effect of the complexity of the polymer on the entropic forces arising in
the confined geometry.
We  investigated the behaviour of ring polymer chains such as unknotted
$0_{1}$ or with knots $3_1,6_1,7_1,9_1,10_{124}$ (according to the
standard Alexander-Briggs notation $c_p$) and twist knot with
\cite{KA} $n=20$ (see table~\ref{knotrg}). The $3_1$ knot is the example of one
of the simplest non-trivial topologies also called trefoil
knot. In this notation, the $c$  is the number of crossings, 
whereas the subscript $p$ corresponds to the order
among all knots with the same number of crossings, which can be an arbitrarily assigned number.

{ We calculated the radii of gyration for each knot type using MD simulations at an adsorption temperature of $T=2.30$ \cite{KremmerBinder}.
}
The obtained results are presented in
 table~\ref{knotrg}. One can notice that for the least complicated
polymers like $3_1$, the radius of gyration is the highest, while for the
most complicated ring polymer with the twist knot $n=20$ it has the
lowest value, which is an expected observation.

Additionally, we  performed molecular dynamics simulations of the model and calculated the entropic forces  for different topological structures of ring polymer chains with the excluded volume interaction confined in a slit geometry of two parallel walls where one is attractive and the other one is repulsive. As it is known, the entropic force measures the polymer interaction with the confining walls.
\begin{figure}[!t]
	\begin{center}
		\includegraphics[width=13cm,height=5cm]{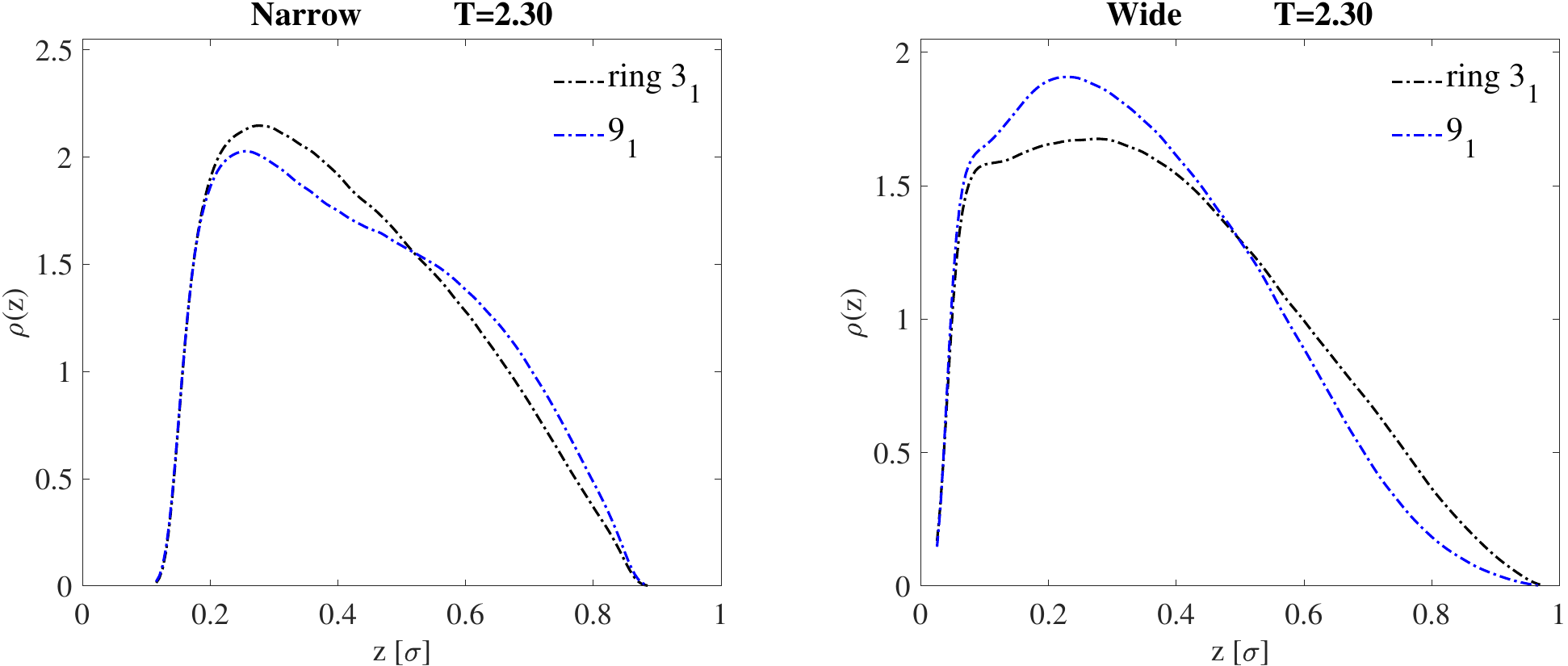}
		\caption{(Colour online) The monomer density profiles for temperature $T=2.30$ in narrow and wide slit for two selected knots: $3_1$ and $9_1$ obtained using molecular dynamic simulations. 
			The slit width $z$ is on the horizontal axis and is scaled to $1$. Narrow slit corresponds to the width of $4.81 \sigma$ ($L< 2 R_g$) and wide slit corresponds to $19.24 \sigma$ ($L>2 R_g$) \cite{KCDJ23}. 	
		}
		\label{fig:mono}
	\end{center}
\end{figure}

The results of simulations are presented in 
figure~\ref{fig:fratio} {and  some analytical results for a wide slit region based on the results obtained in \cite{RU09,UH17} can be found in figure~\ref{fig:force} for qualitative comparison reasons}.
We can observe from figure~\ref{fig:fratio} that as the distance between the walls increases, the entropic forces nonlinearly approach zero.
As it is possible to see from the results presented in figure~\ref{fig:fratio}, the ratio of the entropic forces for ring polymers
decreases with an increase of the slit width and with an increase of the complexity of the polymer topology for a wide slit region. 
{ 
Generally, the real ring polymers with the EVI have a greater absolute value of the forces
than ideal polymer chains, which is an interesting observation from the
theoretical point of view. It suggests that the excluded volume
effects have a significant impact on the properties of polymer
solution. 
As we can see from figure~\ref{fig:mono}, the  monomer density profiles of ring polymers with different topological structures 
have higher values near the attractive wall and decrease nonlinearly in the direction towards the repulsive wall.
Moreover, one can observe that for a narrow slit (left side), the polymers with less complicated structure have a higher density near the attractive wall (at $z = 0$) than more complicated ones. 
The opposite behaviour is observed near the repulsive wall. 
For a wide slit, the relation of densities to polymer complexity across the slit is contrary compared to the narrow case. 
It results in a smaller probability of contact between entangled polymer and this wall. The consequence of this could be 
that for a wide slit region, the degree of force generation due to fluctuations and contact between the walls is significantly reduced.
Thus, it could be understandable that with an increase of the distance between the walls the influence
of the interactions of polymer with walls becomes negligible.
The obtained value of the monomer density is expected to control the force exerted by the polymer on the confining walls. The comparison of figure~\ref{fig:mono} and figure~\ref{fig:fratio} demonstrate a close correlation between the two above mentioned quantities. 
}

{
In figure~\ref{fig:force}, we present the analytical results of entropic force ratios for the given topologies $7_1$,  $9_1$ and $n_{ \rm tw} = 20$ to force for 
linear polymer in a wide slit obtained on the basis of the results presented in~\cite{RU09,UH17}. 
The results suggest that in a wide slit, the forces  decrease with an increase of the complexity of the polymer topology.
These results are in a qualitative agreement with the results obtained by molecular dynamic simulations presented in figure~\ref{fig:fratio} when we look at the right side of the graph, where the wide slit region is located.
}
 Besides, as it is possible to see from figure~\ref{fig:fratio}, each type of ring polymer topologies has its own crossing point
 of the ratio being higher or lower compared to the linear polymer chain, which is shown with the line at the value $1.0$.
It can be seen on the left-hand side of
 figure~\ref{fig:fratio}, that the more complex is the knot, the larger force
ratio it has compared to a linear polymer, whereas on the right-hand side
of figure~\ref{fig:fratio} we observe that  for the wide slits, the
relation is opposite. Thus, it means that the more complex is the polymer, the
smaller force ratio it has.

\section{Conclusions}
{ 
The results of MD simulations are presented in figure~\ref{fig:fratio} for the ratio of the entropic forces. This is a ratio of the entropic forces of real ring polymer chains with the EVI for different topologies to the entropic force of a linear polymer chain. These findings align with the analytical results for a dilute solution of ideal ring polymer chains, as shown in figure~\ref{fig:force}. The analytical results are based on the research presented in references \cite{UH17,RU09}.
}
From the obtained numerical and analytical results, we see that the depletion force, which arises in a dilute 
solution of ring polymer chains with different topological structures confined in a slit geometry with mixed D-N b.c. is repulsive. 
%
Those results are interesting from the point of view of their application to nano- or micro-electromechanical 
systems (NEMS and MEMS), because it allows one to reduce the static friction due to the
depletion repulsion, which prevents them from malfunction.

In nanotechnology, the repulsion between confining walls in a dilute solution of real ring polymer chains with mixed D-N boundary conditions has practical applications. 
It helps preventing the agglomeration of nanoparticles and ceramic nanopowders. Additionally, it reduces the static friction between these particles.
{It can be
used for the creation of a new generation of nano- and micro-electromechanical
devices. Preventing the creation of nanoparticle aglomerates is a very important task in 
the nanotechnology and electronic industry in order to reduce the sizes of manufactured devices.
}

{
Our numerical study indicates that polymers with different topological structures exhibit varied force ratios in relation to the slit size. Such a behavior may play a significant role in biological systems. As depicted in figure~\ref{fig:fratio}, more complex knots (or structures) exert larger force ratios for narrow slits. By contrast, in the region of wider slits, these ratios are smaller compared to the linear polymer case. Simulation results show a nonlinear decrease in the force ratio for the most intricate cases, such as the twisted polymer with $20$ twists, which is more pronounced than in other scenarios.
}

{
The molecular dynamic simulation outcomes suggest that the effect of polymer complexity on the force ratio might arise from the increasing steric effects due to the dense spatial packing of polymers with intricate knots. These complex knots have limited spatial penetration due to fluctuations and are inherently stiffer. This suggests that in wide slits, the interaction with the opposite side may be limited, thereby reducing the force exerted on the opposite wall. The density profiles of polymers with varying knot structures are discussed in \cite{KCDJ23}, where it was demonstrated that the wide slit space penetration diminishes with the knot complexity, see  figure~\ref{fig:mono}. Conversely, in narrow slits, where the distance is short, both complex and simple polymers can directly influence the opposite slit sides \cite{KCDJ23}.
}

In recent papers \cite{DD17,ZZ22}, using the tube model it was shown that the sizes of polymer knots (measured in monomer size units) are influenced by the confinement. Under strong confinement conditions, the knot sizes 
{ 
decrease. A comparison of the free energy required to form a knot in a free space versus under the confinement revealed that the free energy around the knot intersections increases, 
forcing the monomers out of the knot to decrease the overall free energy of the chain. 
}
Our findings show that strongly confined polymers with more complex knots exert larger forces on the walls. 
{ 
Similar to prior findings, it is plausible that the increased forces exerted by knots may also be linked to the enlarged free energies at the knot intersections.
}

{
The dual nature of evolutionarily conserved slit proteins may have broader implications in the organism development. For instance, they have been shown to guide the tracheal development of \textit{Drosophila melanogaster} \cite{ESFXS02}. We highlight a potential mechanism that might switch between interacting and non-interacting walls, achievable by merely altering the polymer's conformation in the medium. In the case of two repulsive walls, the polymer chain stays in the middle of the slit. However, in the case of a slit with mixed b.c., when one wall is attractive and the other one is repulsive, the polymer
prefers to stay near the attractive wall. In the mixed case, changing the topology of the chain can have a much higher impact on the space penetrated by itself due to the thermal fluctuations 
compared to the nonadsorbing case.
}

 In rough ER in cells, such as those in translationally active liver cells, protein translation is a continuous, steady-state process. 
This process may help maintain the rough ER's structure, which is characterized by elongated, flat disks dispersed throughout the cell volume.
Any disruption in this translation stream or changing the interaction properties of polymers, due to adding the knots, could morphologically impact the size of the lumen of ER structure. Such structural changes might result in a potential  risk of the cell malfunction. On the other hand, as was indicated in \cite{DM21}, the ER plays a pivotal role in the axon development and presynaptic differentiation. In other words, such ER structural changes due to polymer switching to the knotted forms, might impact the stability of ER, which could further lead to a potential reduction of neural network development.

{
In summary, ring polymer solutions under mixed boundary conditions exhibit distinct behaviours compared to those in homogenous environments. Given the vast range of conditions in biological systems, introducing these interactions from an entropic perspective could provide deeper insights into cellular-scale biological processes. Simplifying these intricate systems might involve employing coarse-grained methods or effective fields to incorporate entropic or depletive effects, complementing classical approaches. }
It might help to improve the understanding of the models such as the protein transport or behaviour inside the ER.
{ Moreover, acknowledging these interactions could be pivotal in comprehending the neuroplasticity aspects, specifically presynaptic differentiation and axon development, as indicated in~\cite{DM21}. Additionally, this knowledge has potential applications in devising new materials for drug delivery systems, such as micelles, whose behavior might deviate due to entropic or depletive factors compared to standard lab conditions. Overall, considering entropic or depletion interactions in biological systems can enrich our comprehension of these intricate processes.}

Ambivalent nature of evolutionary conserved slit proteins might have wider importance in organism
development. 
It was shown that it can guide tracheal development of \textit{Drosophila melanogaster} \cite{ESFXS02}.
We show potentially one of the mechanisms, which could lead to switching between interacting and noninteracting walls, which could be done by just changing the conformation of 
the polymer in the medium, which results in 
smaller or larger space penetrated by thermally fluctuating polymer that is adsorbed on one of the sides of the slit.
{
The polymer systems in mixed interaction environment can present an interesting behaviour from
theoretical and numerical perspective due to thermal fluctuations and some spatial constraints, which in mesoscopic or microscopic scale may  exhibit new phenomena.  This is just
like mentioned earlier in rough ER of cells in general and liver cells in particular. Such interactions due to a constant stream of translational process of proteins belong to the system of one adsorbed
and another repelled polymer. This system can be imagined to be in a state resembling the steady state, which could help in keeping the structure of rough endoplasmic reticulum. The ER  has long and flat disks that are in the bulk of the cell volume.
Any deviation in such stream of polymer translation could have morphological consequences in the cells structure and thus increase the risk of cell malfunction. Like
it was shown in \cite{DM21},  the ER plays an important role in axon development and presynaptic differentiation. Moreover, if we assume that rough ER width is around 20--30~nm, then switching
conformation by entangling it could switch between interacting and non-interacting states with two walls at the same time.
}

\lastpage

\ukrainianpart

\title{Ентропійна сила в розведеному розчині справжніх ланцюгів кільцевих полімерів з різними топологічними структурами
	в щілині двох паралельних стінок зі змішаними граничними умовами
	}
\author[П. Кутерба, З. Данель, В. Янке]{П. Кутерба\refaddr{label1,label2}, 
	З. Данель\refaddr{label3},
	В. Янке\refaddr{label4}}
\addresses{
	\addr{label1} Інститут теоретичної  фізики, Ягеллонський університет,  вул. проф. С. Лоясєвіча 11, 30--348 Краків, Польща
	\addr{label2} Факультет медицини, Медичний коледж Ягеллонського університету, вул. св. Лазаря 16, 31--530 Краків, Польща
	\addr{label3} Факультет матеріалознавства та фізики, Краківський технологічний університет, вул. Підхорунжих  1 , 30--084 Краків, Польща
	\addr{label4} Інститут теоретичної фізики, Лейпцигський університет, IPF 231101, 04081 Лейпциг, Німеччина
}

\makeukrtitle

\begin{abstract}
	Моделювання методом молекулярної динаміки було використано для розрахунку радіуса гірації справжніх ланцюжків із кільцевих полімерів  з
	різними топологічними структурами, що містять 360 мономерів.
	Автори зосереджуються на ентропійній силі, з якою розведений розчин ланцюжків кільцевого полімеру
	різних топологічних структур з взаємодією виключеного об'єму
	(EVI) діє у хорошому розчиннику на обмежуючі паралельні стінки з щілинною геометрією. Розглядаються
	змішані граничні умови, коли одна стінка є відштовхуючою, 
	а інша – на границі адсорбції.
	Отримані результати молекулярної динаміки для  області широкої щілини
	показують якісне узгодження з попередніми аналітичними результатами
	для ідеальних кільцевих полімерів. 
	Ці результати можуть мати цікаве потенційне
	застосування у матеріалознавстві та розширювати наше розуміння
	деяких біологічних процесів, про які йдеться у статті. Крім того, це
	може бути застосовано в мікро- та наноелектромеханічних пристроях
	для зниження ефекту статичного тертя.
	\keywords м'яка речовина, термодинаміка інтерфейсів та поверхонь, поведінка полімерів, фазовий перехід, молекулярна динаміка
	водневі зв'язки
\end{abstract}
\end{document}